\documentclass[aps,preprint,a4paper,amsmath,amssymb,floatfix]{revtex4}
\usepackage{exscale}
\usepackage{graphicx}
\usepackage[colorlinks, bookmarks=true]{hyperref}

\newcommand{\sech}{\mathrm{sech}}
\newcommand{\Real}{\mathrm{Re}}
\newcommand{\Imag}{\mathrm{Im}}

\newcommand{\deriv}[2]{\frac{d#1}{d#2}}
\newcommand{\nderiv}[3]{\frac{d^#3#1}{{d#2}^#3}}
\newcommand{\pderiv}[2]{\frac{\partial#1}{\partial#2}}
\newcommand{\fnlderiv}[2]{\frac{\delta#1}{\delta#2}}

\newenvironment{bsmatrix}{\left[\begin{smallmatrix}}{\end{smallmatrix}\right]}

\newcommand{\BigFig}[1]{\parbox{12pt}{\Huge #1}}
\newcommand{\BigZero}{\BigFig{0}}

\newcommand{\ket}[1]{\left|#1\right>}
\newcommand{\bra}[1]{\left<#1\right|}
\newcommand{\eigs}[1]{\left|\varphi_{#1}\right>}
\newcommand{\bracketsO}[3]{\left<#1 \left|#2 \right|#3 \right>}
\newcommand{\bracketsObiggm}[3]{\left<#1\! \biggm|\!#2 \!\biggm|\!#3 \right>}
\newcommand{\bracketsbiggm}[2]{\left<#1 \biggm| #2\right>}
\newcommand{\exval}[1]{\left<#1 \right>}
\newcommand{\operator}[1]{\mathbf{\hat#1}}
\newcommand{\commut}[2]{\left[#1, #2\right]}

\newcommand{\ie}{i.\,e.\ }

\newcommand{\epsw}{\bar{\epsilon}(\omega)}

\newcommand{\muw}{\overline{\exval{\operator{\mu}}}(\omega)}
\newcommand{\Ow}{\overline{\exval{\operator{O}}}(\omega)}
\newcommand{\Oaw}{\overline{\exval{\operator{O}_a}}(\omega)}

\newcommand{\feps}{f_{\epsilon}(\omega)}
\newcommand{\tfeps}{\tilde{f}_{\epsilon}(\omega)}
\newcommand{\fmu}{f_{\mu}(\omega)}
\newcommand{\tfmu}{\tilde{f}_{\mu}(\omega)}
\newcommand{\fO}{f_{O}(\omega)}
\newcommand{\tfO}{\tilde{f}_{O}(\omega)}

\newcommand{\evmu}{\exval{\operator{\mu}}\!(t)}

\begin{document}
\title{Optimal control theory of harmonic generation}
\author{Ido Schaefer and Ronnie Kosloff}
%\address{The Fritz Haber Research Center for Molecular Dynamics\\
\affiliation{The Fritz Haber Research Center for Molecular Dynamics\\
The Institute of Chemistry\\
The Hebrew University of Jerusalem\\
Jerusalem 91904, Israel}
\email{ido.schaefer@mail.huji.ac.il}
\email{ronnie@fh.huji.ac.il}
\date{\today}
\begin{abstract}
Coherent control of harmonic generation was studied theoretically. A specific harmonic order was targeted.
An optimal control theory was employed to find the driving field where restrictions were imposed on the frequency band. 
Additional restrictions were added to suppress undesired 
outcomes such as ionization and dissociation. The method was formulated in the 
frequency domain. An update procedure for the field based on relaxation was employed. 
The method was tested on several 
examples demonstrating generation of high frequencies from a driving field with a restricted frequency 
band.
\end{abstract}
\maketitle

\section{Introduction}
\label{sec:intro}

When high intensity light is irradiated on an atomic or molecular gas, 
higher frequencies are generated \cite{burnet77,ferray88}.
This phenomenon is known as high harmonic generation \cite{corkum93,lewenstein94,agostini99}.
The current approach to attosecond pulse generation \cite{mauritsson06} is based on this phenomenon. 
Significant effort has therefore been devoted to optimizing the process of harmonic generation \cite{agostini99,itanani05,lewnstein98}. 
Optimizing by control of the phase and amplitude of the incident light suggests coherent control \cite{HHGcol}.

The present paper addresses theoretically the issue of an optimal control strategy for harmonic generation. 
The target of optimization is the system's dipole operator. To be a source of radiation the acceleration of the dipole operator should oscillate in the 
frequency which is much higher than the driving field frequency.
The idea is to exploit the significant theoretical development in optimal control theory (OCT) \cite{Rabitz,k67,k193,gross07,Degani}.
The hope is that the optimized pulses will unravel specific mechanisms of harmonic generation.

OCT for harmonic generation poses two significant challenges: 
\begin{itemize}
	\item{A constraint on the bandwidth of the incident control pulse has to be imposed;}
	\item{The target is designated in the frequency domain while OCT is typically formulated in the time domain.}
\end{itemize}
Several suggestions for dealing with the first issue appear in the literature in a more general context \cite{gross07, Degani, Skinner, gollub08, gollub10, sugny09, brown09, motzoi11}.
It has been attempted to deal with the second issue by the means of the general OCT formulation of time-dependent targets \cite{serban05, gross07}. However, no results from this approach have been reported.

In the present study, the two challenges are overcome by the formulation of the control problem in the frequency domain, replacing the formulation in the time domain. In this approach, the frequency requirements of the problem are expressed in a natural and direct way. A simple and effective optimization procedure, suitable for the new formulation, is suggested.

\section{OCT of time dependent problems}
\label{sec:TD}
We first review the formulation of optimal control theory for time dependent problems.
Let us denote the time-dependent state of the system by $\ket{\psi(t)}$, the drift Hamiltonian by 
$\operator{H}_0$, and the driving field by $\epsilon(t)$. The dynamics of the system is governed by the 
time-dependent Schr\"odinger equation, under a given initial condition:
\begin{equation}
	\pderiv{\ket{\psi(t)}}{t} = -i\operator{H}(t)\ket{\psi(t)}, ~~~~~~ 	\ket{\psi(0)}  = \ket{\psi_0} \label{eq:Schrodinger}\\	
\end{equation}
where $\operator{H}(t) = \operator{H}_0 - \operator{\mu}\epsilon(t)$.
(Atomic units are used throughout, so we set: $\hbar=1$.) 
The optimization functional for time-dependent targets becomes (see~\cite{serban05, gross07, k236, thesis}):
\begin{align}
	& J \equiv J_{max} + J_{bound} + J_{penal} + J_{con} \label{eq:JTD}\\
	& J_{max} \equiv \int_0^T w(t)\bracketsO{\psi(t)}{\operator{O}(t)}{\psi(t)} \,dt & w(t)\geq 0, \quad \int_0^T w(t)\,dt = 1 \label{eq:JmaxTD}\\
	& J_{bound} \equiv \kappa\bracketsO{\psi(T)}{\operator{O}(T)}{\psi(T)} & \kappa \geq 0 \label{eq:JboundTD}\\
	& J_{penal} \equiv -\alpha\int_0^T\epsilon^2(t)\,dt & \alpha>0 \label{eq:JpenalTD}\\
	& J_{con} \equiv -2\Real{\int_0^T\bracketsO{\chi(t)}{\pderiv{}{t}+i\operator H(t)}{\psi(t)}\,dt} \label{eq:JconTD}
\end{align}
where $\operator{O}(t)$ is the time-dependent target operator, $\ket{\chi(t)}$ is a Lagrange-multiplier function, and $T$ is the final time. $J_{max}$ represents the target to be maximized. $J_{bound}$ is a boundary term. The inclusion of this term prevents boundary problems (see~\cite[Section~2.2]{thesis}). $J_{penal}$ is a penalty term on the intensity of $\epsilon(t)$. $J_{con}$ represents the constraint on the dynamics of the system --- the time-dependent Schr\"odinger equation. 

The resulting Euler-Lagrange equations become:
\begin{align}
	&\pderiv{\ket{\psi(t)}}{t} = -i\operator{H}(t)\ket{\psi(t)}, & 	\ket{\psi(0)} = \ket{\psi_0} \label{eq:ELTDpsi} \\ 
	&\pderiv{\ket{\chi(t)}}{t} = -i\operator{H}(t)\ket{\chi(t)} - w(t)\operator{O}(t)\ket{\psi(t)}, & 	\ket{\chi(T)} = \kappa\operator{O}(T)\ket{\psi(T)} \label{eq:ELTDchi} \\ 
	&\operator{H}(t) = \operator{H}_0 - \operator{\mu}\epsilon(t) \nonumber \\
	&\epsilon(t) = -\frac{\Imag{\bracketsO{\chi(t)}{\operator{\mu}}{\psi(t)}}}{\alpha} \label{eq:ELTDeps}
\end{align}

This formulation is suitable for targets that are well defined in the time domain. 
For problems with frequency requirements this approach has to be modified.

\section{OCT of harmonic generation}
\label{sec:method}
The harmonic generation problem may be divided into two distinct parts:
\begin{enumerate}
	\item The driving field spectrum has to be restricted to the frequency range available from the source;
	\item The intensity of the emission of the system in the desired portion of the spectrum has to be maximized.
\end{enumerate}
These two parts will be treated separately in the next two subsections. 

\subsection{Restriction of the driving field spectrum}\label{ssec:res}

The task of restricting the driving field spectrum is of considerable importance in OCT; 
the reason is that most of the computed fields turn out to be too oscillatory to be produced experimentally. 
This problem may be overcome by limiting the spectrum of the field to sufficiently low frequencies. 

Several approaches for achieving this goal have been proposed. In~\cite{gross07}, a spectral filtration of the driving field is performed in each iteration of the Krotov algorithm. This approach leads to a non-monotonic convergence of the optimization procedure. In~\cite{Degani}, a two-dimensional penalty term is introduced in order to control the spectral properties of the driving field. This approach might lead to numerical instabilities or to non-monotonic convergence of the optimization procedure (see~\cite[Section~3.2.1]{thesis}). In~\cite{Skinner}, the problem of optimization of a general driving field function is replaced by the optimization of the coefficients of a list of frequency terms.

In the present approach the restriction on the field spectrum is achieved by placing a penalty function on the undesirable frequency components of the field. 
The regular $J_{penal}$ from Eq.~(\ref{eq:JpenalTD}) is replaced by a penalty term formulated in the frequency domain. 
In~\cite[Section~3.1.2]{thesis}, it is shown that there is a close relationship between this formulation and the methods presented in \cite{gross07, Degani, Skinner}. 

The cosine transform is employed as a spectral tool. Other spectral transforms (\ie, the Fourier transform or the sine transform) could be used as well. 
For a typical signal, a cosine series is known to converge faster than a Fourier series or a sine series (see~\cite[Section~3.1.1]{thesis}).

The operation of the cosine transform on an arbitrary function $g(t)$ is denoted by the symbol $\mathcal{C}$, and the transformed function by $\bar{g}(\omega)$:
\begin{equation}\label{eq:costrans}
	\bar{g}(\omega) \equiv \mathcal{C}[g(t)] \equiv \sqrt{\frac{2}{\pi}}\int_0^\infty g(t)\cos(\omega t)\,dt
\end{equation}
The inverse cosine transform will be denoted by $\mathcal{C}^{-1}$:
\begin{equation}\label{eq:invcostrans}
	\mathcal{C}^{-1}[\bar{g}(\omega)] \equiv \sqrt{\frac{2}{\pi}}\int_0^\infty \bar{g}(\omega)\cos(\omega t)\,d\omega = g(t)
\end{equation}

The driving field in the frequency domain is defined by a finite time cosine transform:
\begin{equation}\label{eq:epsw}
	\epsw = \sqrt{\frac{2}{\pi}}\int_0^T \epsilon(t)\cos(\omega t)\,dt
\end{equation}

The maximal cutoff driving frequency is denoted by $\Omega$. The penalty term from Eq.~(\ref{eq:JpenalTD}) is modified to:
\begin{equation}\label{eq:Jpenalw}
	J_{penal} \equiv -\alpha\int_0^\Omega\frac{1}{\feps}\bar{\epsilon}^2(\omega)\,d\omega \qquad\qquad \alpha>0  
\end{equation}
where $\feps$ is an adjustable function, which satisfies the conditions:
\begin{equation}\label{eq:feps}
	\int_0^\Omega \feps\, d\omega = 1, \qquad\qquad \feps > 0 
\end{equation}
$\alpha$ determines the cost of large fields, Cf.~Eq.~(\ref{eq:JpenalTD}). $\feps$ is chosen to have small values for undesirable frequencies and regular values for the allowed frequency region. It may be interpreted as a filter function, as can be seen in Section~\ref{ssec:EL}\@. An additional envelope shape can be forced on the profile of the driving field spectrum by choosing an appropriate filter function. A complete filtration of undesirable frequencies is achieved in the limit \text{$\feps\longrightarrow 0$} for undesirable $\omega$ values. For practical purposes, $\feps$ may be set to $0$ for these values.

\subsection{Optimization functional for harmonic generation}
\label{ssec:HGfun}
The field emitted by the system consists of the frequencies contained in the spectrum of the dipole expectation. 
In order to maximize the emission in a desired frequency region, the amplitude of the dipole moment oscillations in this frequency region is maximized. 
Thus, the physical quantity of interest is the dipole moment expectation value:
\begin{equation}\label{eq:evmu}
	\evmu = \bracketsO{\psi(t)}{\operator{\mu}}{\psi(t)}
\end{equation}
The dipole spectrum becomes:
\begin{equation}\label{eq:muw}
	\muw = \mathcal{C}[\evmu]=\sqrt{\frac{2}{\pi}}\int_0^T \evmu\cos(\omega t)\,dt
\end{equation}
To maximize the emission in the desired region of the spectrum, the following functional is chosen:
\begin{equation}\label{eq:Jmaxw}
	J_{max} \equiv \frac{1}{2}\lambda\int_0^\Omega \fmu\overline{\exval{\operator{\mu}}}^2(\omega)\,d\omega \qquad\qquad \lambda>0
\end{equation}
where $\fmu$ satisfies the conditions:
\begin{equation}\label{eq:fmu}
	\int_0^\Omega \fmu\, d\omega = 1, \qquad\qquad \fmu\geq 0 
\end{equation}
$\lambda$ is an adjustable coefficient, which determines the relative importance of $J_{max}$. $\lambda$ is redundant with $\alpha$ but from numerical stability considerations it is useful to vary it independently.  $\fmu$ is a filter function, which has pronounced values in the frequency region of interest.

Eq.~(\ref{eq:Jmaxw}) can be generalized to an arbitrary Hermitian operator $\operator{O}$:
\begin{align}
	& J_{max} \equiv \frac{1}{2}\lambda\int_0^\Omega \fO\overline{\exval{\operator{O}}}^2(\omega)\,d\omega & \lambda>0 \label{eq:JmaxwO}\\
	& \Ow = \mathcal{C}\left[\exval{\operator{O}}(t)\right] \label{eq:Ow} \\
	& \int_0^\Omega \fO\, d\omega = 1, & \fO\geq 0 \label{eq:fO}
\end{align}
One possible application of this generalization is in the case when it is desired to maximize emission with polarization other than that of the control field. For instance, if the control field is $x$ polarized and we require emission of $y$ polarized field, $\operator{\mu}$ in $\operator{H}(t)$ is set to be $\operator{\mu}_x$, and \text{$\operator{O} \equiv \operator{\mu}_y$}.

The full maximization functional for the harmonic generation problem becomes:
\begin{align}
	& J \equiv J_{max} + J_{penal} + J_{con} \label{eq:Jw}\\ 
	& J_{max} \equiv \frac{1}{2}\lambda\int_0^\Omega \fO\overline{\exval{\operator{O}}}^2(\omega)\,d\omega \label{eq:Jmaxw2} \\
	& J_{penal} \equiv -\alpha\int_0^\Omega\frac{1}{\feps}\bar{\epsilon}^2(\omega)\,d\omega \label{eq:Jpenalw2}\\
	& J_{con} \equiv -2\Real{\int_0^T\bracketsO{\chi(t)}{\pderiv{}{t}+i\operator H(t)}{\psi(t)}\,dt} \label{eq:Jconw}
\end{align}

\subsection{The Euler-Lagrange equations for harmonic generation}
\label{ssec:EL}
We choose the functional derivative of the objective in the frequency domain:
\begin{equation}\label{eq:dJdepsw}
	\fnlderiv{J}{\bar{\epsilon}(\omega)} = 0
\end{equation}
The resulting Euler-Lagrange equations become: %(see~\autoref{app:der}):
\begin{align}
	&\pderiv{\ket{\psi(t)}}{t} = -i\operator{H}(t)\ket{\psi(t)}, & 	\ket{\psi(0)} = \ket{\psi_0} \label{eq:ELpsiw} \\ 
	&\pderiv{\ket{\chi(t)}}{t} = -i\operator{H}(t)\ket{\chi(t)} - \lambda\,\mathcal{C}^{-1}\left[\fO\Ow\right]\operator{O}\ket{\psi(t)}, & \ket{\chi(T)} = 0 \label{eq:ELchiw} 
\end{align}
where $\operator{H}(t) = \operator{H}_0 - \operator{\mu}\epsilon(t)$ and
\begin{align}
	& \epsw = \feps\mathcal{C}[\eta(t)], \qquad\qquad \eta(t) \equiv -\frac{\Imag{\bracketsO{\chi(t)}{\operator{\mu}}{\psi(t)}}}{\alpha}\label{eq:ELepsw} \\
	& \epsilon(t) = \mathcal{C}^{-1}[\epsw] = \mathcal{C}^{-1}\left\lbrace \feps\mathcal{C}[\eta(t)]\right\rbrace \label{eq:ELepst}
\end{align}
Note that the expression for $\eta(t)$ is the same as that for $\epsilon(t)$ in Eq.~(\ref{eq:ELTDeps}). $\epsilon(t)$ in Eq.~(\ref{eq:ELepst}) can be 
interpreted as the filtered field from the regular control problems, where $\feps$ plays the role of a filter function. 
A comparison between Eq.~(\ref{eq:ELTDchi}) and Eq.~(\ref{eq:ELchiw}) leads to a similar interpretation of the inhomogeneous term in 
Eq.~(\ref{eq:ELchiw}) (Cf.~\cite[Section 3.3.1]{thesis}).

It is convenient to avoid normalizing $\feps$ and $\fO$, and substitute them with:
\begin{align}
	& \tfeps \equiv \frac{\feps}{\alpha} \label{eq:tfeps} \\
	& \tfO \equiv \lambda\fO \label{eq:tfO}
\end{align}

\subsection{Optional modifications for the optimization problem}
\label{ssec:mod}
\subsubsection{Prevention of dissociation}
\label{sssec:dis}
Typically, when strong driving fields are employed the system dissociates or ionizes. This phenomenon can be avoided by restricting the system to localize in an ``allowed'' subspace of the Hilbert space \cite{k236}. For example, eliminating access to all eigenstates with energies above a threshold energy. Another option is to restrict the state vector to regions of space far from
the threshold of the potential well. A similar idea is to restrict the dynamics to the allowed momentum values.

In order to restrict the system to the allowed subspace two modifications in the maximization functional $J$ are employed:
\begin{enumerate}
	\item $J_{max}$ is modified to include contribution only from the allowed states;
	\item A penalty term on the forbidden states is added to $J$.
\end{enumerate}

The first modification is achieved by the replacement of the expectation $\exval{\operator{O}}(t)$ in Eq.~(\ref{eq:Ow}) by the expression:
\[
	\bracketsO{\operator{P}_{a}\psi(t)}{\operator O}{\operator{P}_{a}\psi(t)}
\]
where $\operator{P}_{a}$ is the projection operator onto the allowed subspace. For instance, if all energies above the threshold $E_L$ are restricted, then:
\begin{equation}\label{eq:projal}
	\operator{P}_{a} \equiv \sum_{n=0}^L\ket{\varphi_n}\bra{\varphi_n}
\end{equation}
If the system is restricted in the $x$ space to remain in the interval \text{$[x_{min}, x_{max}]$}, then:
\begin{equation}\label{eq:projalx}
	\operator{P}_{a} \equiv \int_{x_{min}}^{x_{max}}\ket{x}\bra{x}\,dx
\end{equation}

If a smooth filtration of states is desired, $\operator{P}_{a}$ may be generalized to a weighted projection operator, which will be denoted as $\operator{P}_a^s$. For instance, $\operator{P}_{a}$ from Eq.~\eqref{eq:projal} is modified to:
\begin{equation}\label{eq:projals}
	\operator{P}_{a}^s \equiv \sum_{n=0}^{N-1} s_n\ket{\varphi_n}\bra{\varphi_n} \qquad\qquad 0\leq s_n\leq 1	
\end{equation}
where $s_n$ decreases gradually from $1$ to $0$ near the threshold. $\operator{P}_{a}$ from Eq.~\eqref{eq:projalx} is modified to:
\begin{equation}\label{eq:projalxs}
	\operator{P}_{a}^s \equiv \int s(x)\ket{x}\bra{x}\,dx = s\left(\operator{X}\right) \qquad\qquad 0\leq s(x)\leq 1
\end{equation}
where $s(x)$ decays to $0$ near the boundaries of the allowed $x$ interval.
	
The resulting modified $J_{max}$ becomes:
\begin{align}
	& J_{max} \equiv \frac{1}{2}\int_0^\Omega \tfO\overline{\exval{\operator{O}_a}}^2(\omega)\,d\omega \label{eq:Jmaxal}\\
	& \operator{O}_a = \operator{P}_{a}^s\operator O\operator{P}_{a}^s \nonumber
\end{align}

The second modification is the addition of the following penalty term to $J$ (as suggested in~\cite{k236}):
\begin{equation}\label{eq:Jforb}
	J_{forb} \equiv -\gamma\int_0^T \bracketsO{\psi(t)}{\operator{P}_f}{\psi(t)}\,dt \qquad\qquad \gamma>0
\end{equation}
where $\operator{P}_f$ is the projection onto the forbidden subspace and $\gamma$ is the penalty factor of the forbidden subspace.

It is possible to achieve a smooth filtration of states by using a state-dependent penalty factor. For instance, in the energy space Eq.~\eqref{eq:Jforb} is generalized to:
\begin{align}
	& J_{forb} \equiv -\int_0^T \bracketsO{\psi(t)}{\operator{P}_f^\gamma}{\psi(t)}\,dt \label{eq:Jforb2} \\
	& \operator{P}_f^\gamma \equiv \sum_{n=0}^{N-1}\gamma_n\ket{\varphi_n}\bra{\varphi_n} & \gamma_n\geq 0 \label{eq:Pfg}
\end{align}
$\operator{P}_f^\gamma$ is a skewed projection. $\gamma_n$ is the penalty factor of the state $\eigs{n}$. It should be $0$ for the allowed energy domain and increase gradually with $n$ near the threshold energy. In the $x$ space the skewed projection is:
\begin{equation}\label{eq:pfgx}
	\operator{P}_{f}^\gamma \equiv \int \gamma(x)\ket{x}\bra{x}\,dx = \gamma\left(\operator{X}\right) \qquad\qquad \gamma(x)\geq 0
\end{equation}
where $\gamma(x)$ increases gradually in the forbidden $x$ regions. A smooth penalization is recommended, in order to decrease the difficulty in the optimization process.

When the Hilbert space is very large, it becomes impractical to compute all the eigenstates. Nevertheless, a restriction of the allowed subspace in the energy space is still possible. $s_n$ and $\gamma_n$ should be defined as functions of the energy, \ie:
\[
	s_n=s(E_n) \qquad\qquad \gamma_n=\gamma(E_n)
\]
Then we have:
\begin{equation}
	\operator{P}_a^s = s\left(\operator{H}_0\right) \qquad\qquad \operator{P}_f^\gamma = \gamma\left(\operator{H}_0\right)
\end{equation}
$s\left(\operator{H}_0\right)\ket{\psi(t)}$ and $\gamma\left(\operator{H}_0\right)\ket{\psi(t)}$ can be approximated using standard methods.

When \text{$\operator{P}_{a}^s = \operator{I}$} and \text{$\operator{P}_f^\gamma = \operator{0}$}, $J$ reduces to Eq.~(\ref{eq:Jw}).

After inserting these changes in $J$, the equation for $\ket{\chi(t)}$ Eq.~(\ref{eq:ELchiw}) is modified in the following way: %(see~\autoref{app:der}):
\begin{equation} \label{eq:ELchiforb}
	\pderiv{\ket{\chi(t)}}{t} = -i\operator{H}(t)\ket{\chi(t)} - \left\lbrace\mathcal{C}^{-1}\left[\tfO\overline{\exval{\operator{O}_a}}(\omega)\right]\operator{O}_a - \operator{P}_f^\gamma\right\rbrace\ket{\psi(t)}
\end{equation}

\subsubsection{Prevention of boundary effects}

In practice, a finite time spectral transform is approximated by a discrete series. 
Care should be taken on possible boundary effects, otherwise noise (``ringing'') throughout the spectral representation of the signal is generated. For a cosine representation this effect is relatively small. If necessary, it is possible to reduce this phenomenon by trying to enforce the boundary conditions.
The appropriate boundary conditions for a cosine series representation of $\exval{\operator{O}}(t)$ are:
\begin{equation}
	\deriv{\exval{\operator{O}}(0)}{t}=0 \qquad\qquad \deriv{\exval{\operator{O}}(T)}{t}=0 \label{eq:condT}
\end{equation}
Usually, the condition at $t=0$ is automatically satisfied because the initial state is typically chosen to be the ground state. The condition at $t=T$ may be enforced by an addition of the following penalty term to $J$:
\begin{equation}\label{eq:Jbring}
	J_{bound} \equiv -\frac{1}{2}\kappa\left[\deriv{\exval{\operator{O}}(T)}{t}\right]^2 \qquad\qquad \kappa \geq 0
\end{equation}
$\kappa$ is an adjustable parameter. %, which has the role of a penalty factor.
When \text{$\kappa=0$}, $J$ reduces to Eq.~(\ref{eq:Jw}).

In the special case that \text{$\commut{\operator{\mu}}{\operator{O}}=\operator{0}$}, the insertion of $J_{bound}$ results in a relatively simple modification of the Euler-Lagrange equations. The natural boundary condition for $\ket{\chi(t)}$ in Eq.~(\ref{eq:ELchiw}) is replaced by: %(see~\autoref{app:der}):
\begin{equation}\label{eq:chiTring}
	\ket{\chi(T)} = \kappa\exval{\left[\operator{H}_0, \operator{O}\right]}(T) \left[\operator{H}_0, \operator{O}\right]\ket{\psi(T)}
\end{equation}
The general case is more complex and will not be discussed here.

The derivation of the Euler-Lagrange equations is presented in Appendix~\ref{app:der}.

\subsection{Optimization procedure}\label{ssec:opt}
For current optimization posed both in time and frequency, the well established optimization procedures do not converge monotonically or converge very slowly.
We therefore employed a more direct \emph{relaxation method} to update the driving field from iteration to iteration. 
In the present context the method consists of the following update rule for the driving field:
\begin{align}
	& \bar{\epsilon}^{new}(\omega) = K\bar{\epsilon}^{EL}(\omega) + (1-K)\bar{\epsilon}^{old}(\omega) \qquad\qquad 0<K\leq 1 \label{eq:relaxw} \\
	& \bar\epsilon^{EL}(\omega) \equiv \tfeps\mathcal{C}\left[-\Imag{\bracketsO{\chi(t)}{\operator{\mu}}{\psi(t)}}\biggm|_{\epsw=\bar\epsilon^{old}(\omega)}\right] \label{eq:epsELw}
\end{align}
The updated field is a mixture of the previous field and the field computed from the Euler-Lagrange equation Eq.~(\ref{eq:ELepsw}), 
using the previous field for the computation of $\ket{\chi(t)}$ and $\ket{\psi(t)}$. $K$ is the mixing parameter, which determines the weights of the two fields. The value of $K$ is decreased when the optimization process progresses.

The scheme for the implementation of the relaxation method becomes:
\begin{enumerate}
	\item Guess a driving field spectrum $\bar\epsilon^{(0)}(\omega)$. \label{pr:eps0}
	\item Set: $\epsilon^{(0)}(t) = \mathcal{C}^{-1}[\bar\epsilon^{(0)}(\omega)]$.
	\item Guess an initial value for $K$. \label{pr:upK}
	\item Propagate $\ket{\psi^{(0)}(t)}$ forward from $t=0$ to $t=T$ according to Eq.~(\ref{eq:ELpsiw}), with $\epsilon^{(0)}(t)$.
	\item Calculate $J^{(0)}$ with $\ket{\psi^{(0)}(t)}$ and $\bar\epsilon^{(0)}(\omega)$.
	\item (k = 0)
	\item Repeat the following steps until convergence:\label{pr:rdoconv}
	\begin{enumerate}
		\item Set $\ket{\chi^{(k)}(T)}$ according to Eq.~(\ref{eq:chiTring}), using $\ket{\psi^{(k)}(T)}$.
		\item Propagate $\ket{\chi^{(k)}(t)}$ backward from $t=T$ to $t=0$ according to Eq.~(\ref{eq:ELchiforb}), with $\epsilon^{(k)}(t)$.
		\item Do the following steps, and repeat while $J^{trial}\leq J^{(k)}$:
		\begin{enumerate}
			\item Set a new field, using Eq.~(\ref{eq:relaxw}):
			\begin{align*}
				&\bar\epsilon^{trial}(\omega) = K\tfeps \mathcal{C} \left[-\Imag\bracketsObiggm{\chi^{(k)}(t)}{\operator{\mu}}{\psi^{(k)}(t)}\right] + (1-K)\bar\epsilon^{(k)}(\omega) \\
				& \epsilon^{trial}(t) = \mathcal{C}^{-1}[\bar{\epsilon}^{trial}(\omega)]
			\end{align*}
			\item Propagate $\ket{\psi^{trial}(t)}$ forward from $t=0$ to $t=T$ according to Eq.~(\ref{eq:ELpsiw}), with $\epsilon^{trial}(t)$.
			\item Calculate $J^{trial}$ with $\ket{\psi^{trial}(t)}$ and $\bar\epsilon^{trial}(\omega)$.
			\item If $J^{trial}\leq J^{(k)}$, then set: $K = K/2$
		\end{enumerate}
		\item Update all the variables:
		\begin{equation*}
			\bar{\epsilon}^{(k+1)}(\omega) = \bar{\epsilon}^{trial}(\omega) \qquad \epsilon^{(k+1)}(t) = \epsilon^{trial}(t) \qquad \ket{\psi^{(k+1)}(t)} = \ket{\psi^{trial}(t)} \qquad J^{(k+1)} = J^{trial} 		
		\end{equation*}
		\item (k = k + 1)
	\end{enumerate}
\end{enumerate}

It can be  shown (see~\cite[Section~3.2.3]{thesis}) that the relaxation method, in the context of quantum-OCT problems, can be considered an approximated second order gradient method (quasi-Newton method), where the Hessian of $J$ is approximated by the Hessian of $J_{penal}$. 

\section{Application}\label{sec:appl}
The new method is demonstrated in four simple harmonic generation examples. %, of increasing complexity.
The propagator for the Schr\"odinger equation is based on a new, efficient and highly accurate algorithm, \cite{k273}.

The convergence condition of the optimization procedure is:
\begin{equation}
	 \frac{\Vert \vec{\bar\epsilon}^{\text{ }new}-\vec{\bar\epsilon}^{\text{ }old} \Vert}{\Vert \vec{\bar\epsilon}^{\text{ }new} \Vert}<\tau \label{eq:tolw}\\
 \end{equation}
where $ \vec{\bar\epsilon} $ is the discrete vector of frequency values on an equidistant $\omega$ grid
and $\tau$ is a tolerance parameter. 
More numerical details may be found in \cite[Appendix B]{thesis}.

The important details of the problems and the computational process are presented in the tables. 
Atomic units are used throughout. The notations in the tables are described in Table~\ref{tab:not}\@. $\Theta(x)$ denotes the Heaviside step function.
The initial state in all problems is chosen to be the ground state, denoted as $\eigs 0$.

\begin{table}
\begin{center}
\renewcommand{\arraystretch}{1.5}
\begin{tabular}{|c||c|}	\hline
	\textbf{Notation} & \textbf{Description} \\ \hline \hline
	$\operator{H}_0$ & unperturbed Hamiltonian \\ \hline
	$\operator{\mu}$ & dipole moment operator \\ \hline
	$\ket{\psi_0}$ & initial state vector \\ \hline
	$T$ & final time \\ \hline
	$\tfeps$ & scaled filter function of the driving field\\ \hline
	$\tfmu$ & scaled filter function of the dipole moment expectation value \\ \hline
	$\bar{\epsilon}^{0}(\omega)$ & initial guess of the field \\ \hline
	$L$ & index of the maximal allowed eigenstate \\ \hline
	$\gamma_n$ & penalty factor of the state $\ket{\varphi_n}$ \\ \hline
	$s(x)$ & projection function onto allowed $x$ regions \\ \hline
	$\gamma(x)$ & penalty function on forbidden $x$ regions \\ \hline
	$K_i$ & initial guess of $K$, for the relaxation method \\ \hline
	$x \text{ domain}$ & domain of the $x$ grid \\ \hline
	$N_{x}$ & number of equidistant points in the $x$ grid \\ \hline
	$\tau$ & tolerance parameter of the optimization process \\ \hline
\end{tabular}
\end{center}
\caption{Description of the notations in the tables}\label{tab:not}
\end{table}

\subsection{Two level system}\label{ssec:TLS}
\begin{figure}
	\centering \includegraphics[width=3in]{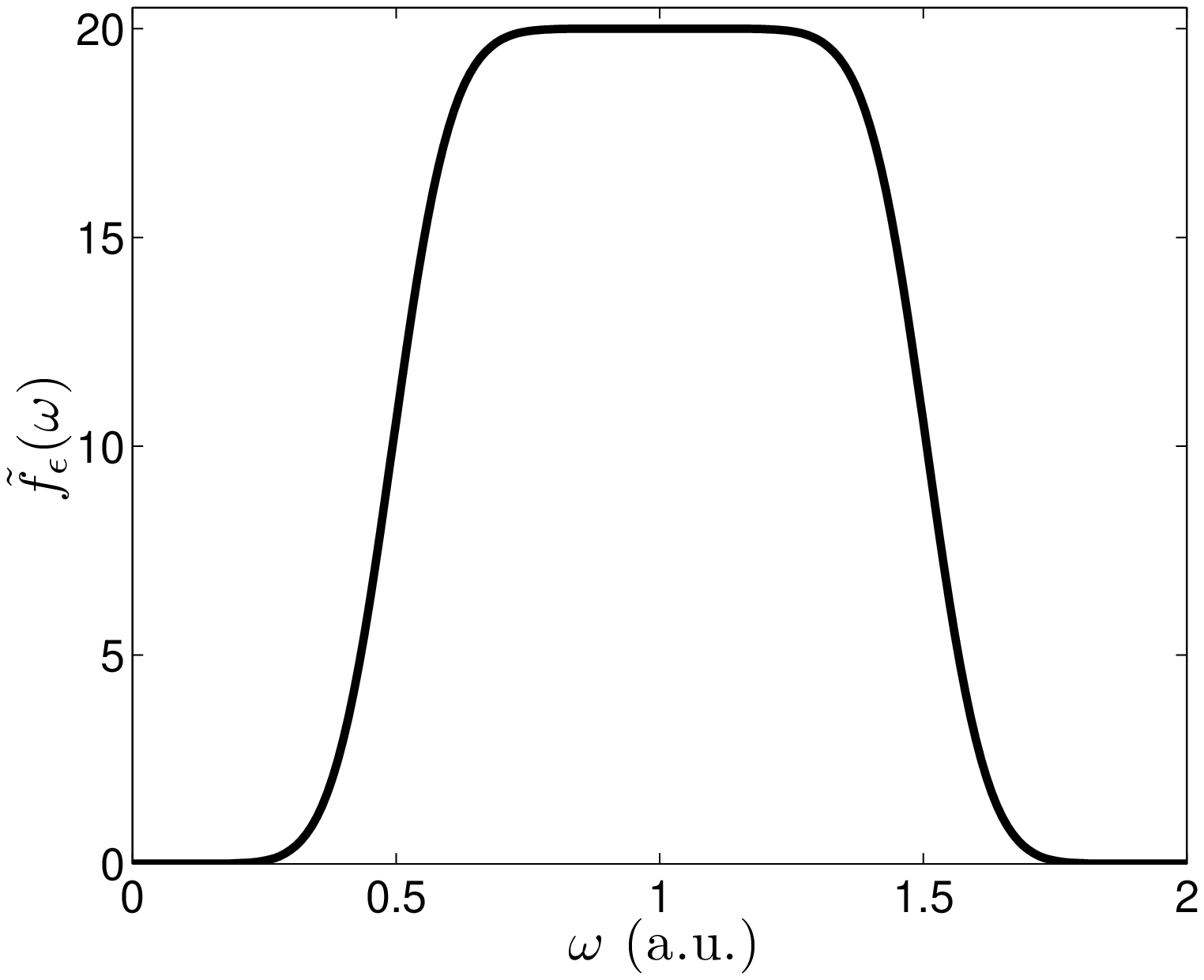}
	\caption{$\tfeps$ of the TLS problem}\label{fig:hat}	
\end{figure}

\begin{figure}
	\centering \includegraphics[width=3in]{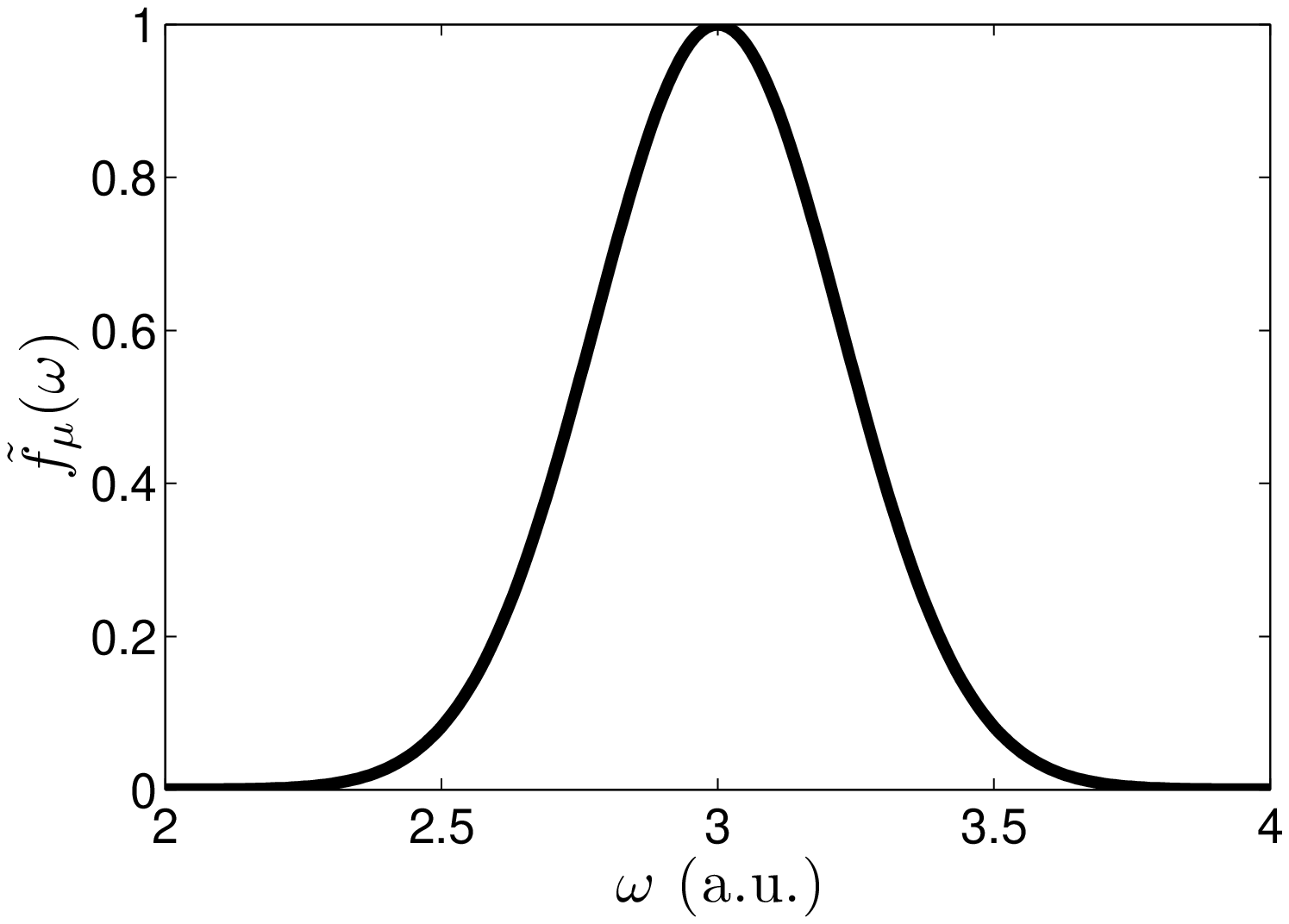}
	\caption{$\tfmu$ of the TLS problem}\label{fig:gauss10}	
\end{figure}

The first problem is tripling the driving frequency by a two level system (TLS). The unperturbed Hamiltonian is:
\begin{equation}\label{eq:TLSH}
	\operator{H}_0 =
	\begin{bmatrix}
		1 & 0 \\
		0 & 4
	\end{bmatrix}
\end{equation}
The dipole moment operator is chosen to be the $x$ Pauli matrix:
\begin{equation}\label{eq:TLSmu}
	\operator{\mu}= \sigma_x =
	\begin{bmatrix}
		0 & 1 \\
		1 & 0
	\end{bmatrix}
\end{equation}
The driving field is restricted to be centered around \text{$\omega=1_{a.u.}$}, by a ``hat'' filter function (Cf.~Fig.~\ref{fig:hat}). 
We require maximization of the emission in the region of the characteristic frequency of the system, \text{$\omega_{1,0}=3_{a.u.}$}. 
A Gaussian function is used for $\tfmu$ (see Fig.~\ref{fig:gauss10}).

The details of the problem are summarised in Table~\ref{tab:TLS}.

\begin{table}
\begin{equation*}
	\renewcommand{\arraystretch}{1.25}
	\begin{array}{|c||c|}
		\hline
		\operator{H}_0 & 
		\begin{bsmatrix}
			1 & 0 \\
			0 & 4
		\end{bsmatrix} \\ \hline
		\operator{\mu} &
		\begin{bsmatrix}
			0 & 1 \\
			1 & 0
		\end{bsmatrix} \\ \hline
		\ket{\psi_0} & 
		\begin{bsmatrix}
			1  \\
			0	
		\end{bsmatrix} \\ \hline
		T & 100 \\ \hline
		\tfeps & 20\,\sech[20(\omega - 1)^4] \\ \hline
		\tfmu & \exp[-10(\omega -3)^2] \\ \hline
		\bar\epsilon^{0}(\omega) & \sech[20(\omega - 1)^4] \\ \hline
		K_i & 0.5 \\ \hline
		\tau & 10^{-3} \\ \hline
	\end{array}
\end{equation*}
\caption{The details of the TLS problem}\label{tab:TLS}
\end{table}

\begin{figure}
	\centering \includegraphics[width=3in]{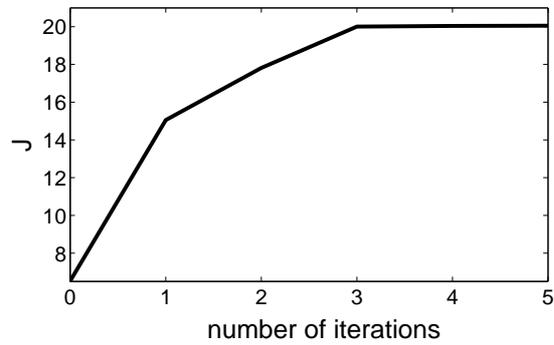}
	\caption{The convergence curve of the TLS problem}\label{fig:TLSHGconv}
\end{figure}

The optimization process converges rapidly to a solution. The convergence curve is shown in Fig.~\ref{fig:TLSHGconv}.

\begin{figure}
	\centering \includegraphics[width=3in]{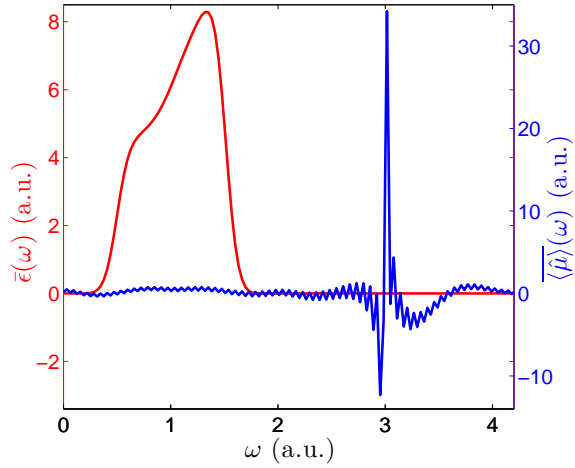}
	\caption{(Color online) The spectra of the driving field (red, gray) and the dipole moment expectation spectra (blue, dark gray), for the TLS problem}\label{fig:TLSepsmu}
\end{figure}

The resulting spectra of the driving field and the dipole moment expectation value are shown in Fig.~\ref{fig:TLSepsmu}. 
$\epsw$ is shown to be successfully restricted to the desired portion of the spectrum. 
The ``hat'' envelope shape is apparent. $\muw$ mainly consists of a large peak at $\omega_{1,0}$, as required.

\subsection{Eleven level system}\label{ssec:11LS}
The second problem is of an eleven level system (11LS). The problem is designed for harmonic generation by a resonance 
mediated absorption mechanism (for example, see~\cite{Amitay}).

The unperturbed Hamiltonian is:
\begin{equation}\label{eq:elLSH}
	\setcounter{MaxMatrixCols}{11}
	\operator{H}_0 = 
	\begin{bmatrix}
		1   &   &   &   &   &   &   &   &   &   & \\
		    &2.1&   &   &   &   &   &   &   &   & \\
		    &   &  3&   &   &   &   &   &   &   & \\
		    &   &   &3.9&   &   &   &   &\BigZero &   & \\
		    &   &   &   &  5&   &   &   &   &   & \\
		    &   &   &   &   &6.1&   &   &   &   & \\
		    &   &   &   &   &   &  7&   &   &   & \\
		    &   &\BigZero &   &   &   &   &8.1&   &   & \\
   		    &   &   &   &   &   &   &   &  9&   & \\
		    &   &   &   &   &   &   &   &   &9.9& \\
		    &   &   &   &   &   &   &   &   &   &11
	\end{bmatrix}
	% E0=[1 2.1 3 3.9 5 6.1 7 8.1 9 9.9 11]
\end{equation}
The dipole moment operator is:
\begin{equation}\label{eq:elLSmu}
	\setcounter{MaxMatrixCols}{11}
	\operator{\mu} =
	\begin{bmatrix}
		   0&  1&   &   &   &   &   &   &   &   & 1\\
		   1&  0&  1&   &   &   &   &   &   &   & \\
		    &  1&  0&  1&   &   &   &   &   &   & \\
		    &   &  1&  0&  1&   &   &   &   &   & \\
		    &   &   &  1&  0&  1&   &   &\BigZero&   & \\
		    &   &   &   &  1&  0&  1&   &   &   & \\
		    &   &\BigZero&   &   &  1&  0&  1&   &   & \\
		    &   &   &   &   &   &  1&  0&  1&   & \\
		    &   &   &   &   &   &   &  1&  0&  1& \\
		    &   &   &   &   &   &   &   &  1&  0& 1\\
		   1&   &   &   &   &   &   &   &   &  1& 0
	\end{bmatrix}
\end{equation}
This $\operator{\mu}$ couples between neighbouring eigenstates, and between the outer eigenstates, $\ket{\varphi_0}$ and $\ket{\varphi_{10}}$. The driving field is restricted so as not to exceed the region of the resonance frequencies of the neighbouring levels, \text{$\omega_{n+1,n}=1_{a.u.}$}. We require an emission in the neighbourhood of the Bohr frequency of the outer energy levels, \text{$\omega_{10,0}=10_{a.u.}$}. $\tfeps$ and $\tfmu$ are chosen to be rectangular functions.

The details of the problem are summarized in Table~\ref{tab:11LS}.

\begin{table}
\begin{equation*}
	\renewcommand{\arraystretch}{1.5}
	\begin{array}{|c||c|}
		\hline
		\operator{H}_0 & \text{Eq.~(\ref{eq:elLSH})} \\ \hline
		\operator{\mu} & \text{Eq.~(\ref{eq:elLSmu})} \\ \hline
		\ket{\psi_0} & \ket{\varphi_0} \\ \hline
		T & 100 \\ \hline
		\tfeps & 50\,\Theta(1.3-\omega) \\ \hline
		\tfmu & \Theta(\omega - 9.9)\,\Theta(10.1-\omega) \\ \hline
		\bar\epsilon^{0}(\omega) & \Theta(1.3-\omega) \\ \hline
		K_i & 1 \\ \hline
		\tau & 10^{-3} \\ \hline
	\end{array}
\end{equation*}
\caption{The details of the 11LS problem}\label{tab:11LS}
\end{table}

\begin{figure}
	\centering \includegraphics[width=3in]{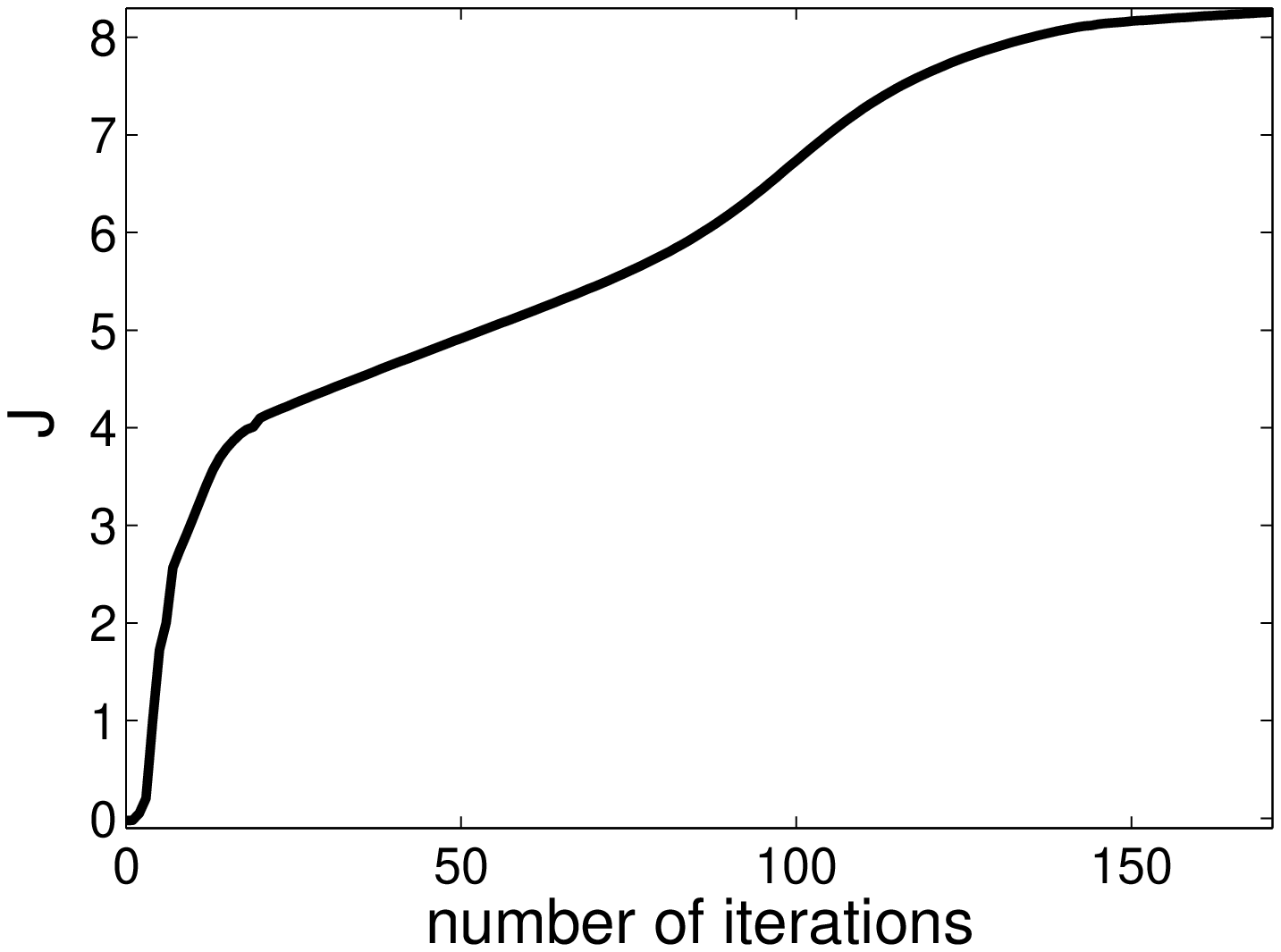}
	\caption{The convergence curve of the 11LS problem}\label{fig:11LSconv}	
\end{figure}

\begin{figure}
	\centering \includegraphics[width=3in]{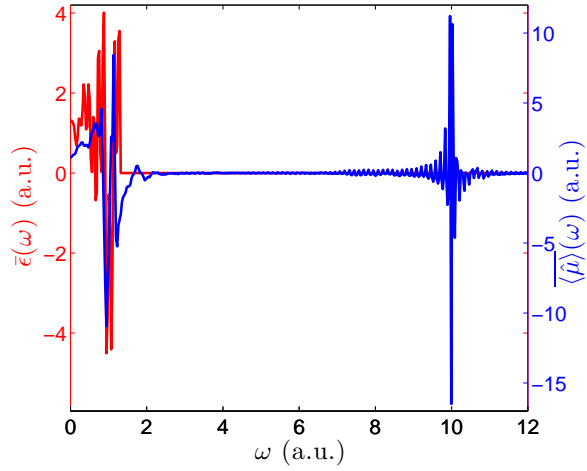}
	\caption{(Color online) The spectra of the driving field (red, gray) and the dipole moment expectation value (blue, dark gray), for the 11LS problem}\label{fig:11LSepsmu}	
\end{figure}

The convergence curve is shown in Fig.~\ref{fig:11LSconv}. The resulting $\epsw$ and $\muw$ are shown in Fig.~\ref{fig:11LSepsmu}. The new method is shown again to be quite efficient in maximizing the emission in the required region. %is shown again to be maximized with quite success.

\subsection{Anharmonic oscillator --- the HCl molecule}\label{ssec:HCl}

In this example an anharmonic oscillator is used to double the incoming frequency. 
The oscillator chosen is an approximation of the H$\,^{35}$Cl molecule (see Appendix~\ref{app:HCl}). 
As in the previous problem, the intended mechanism is of harmonic generation by resonance mediated absorption.

\begin{figure}
	\centering \includegraphics[width=3in]{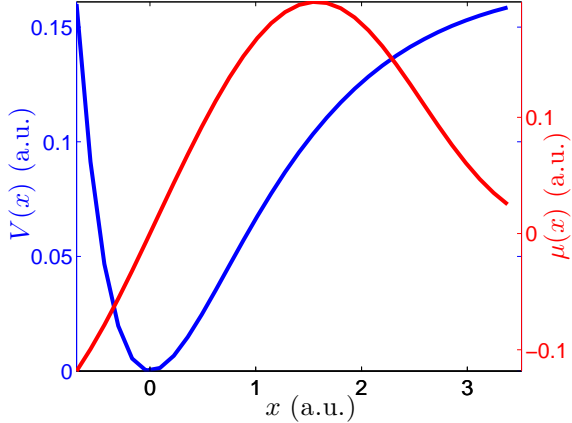}
	\caption{(Color online) The approximated potential (blue, dark gray) and dipole function (red, gray) curves, for the HCl molecule}\label{fig:Vmux}	
\end{figure}

The coordinate of the one-dimensional oscillator is the displacement of the inter-nuclei distance, $r_{H-Cl}$, from the  bottom of the well ($r^*$):
\begin{equation}\label{eq:HClx}
	x = r_{H-Cl} - r^*
\end{equation}
The approximated potential function, $V(x)$, and dipole moment function, $\mu(x)$, are shown in Fig.~\ref{fig:Vmux}.

$\epsw$ is restricted not to exceed much the characteristic frequency of the bottom of the well: %\text{$\omega_0 = 1.35\cdot 10^{-2}_{a.u.}$}.
\begin{equation*}%\label{eq:HClw0}
	\omega_0 = {1.35\cdot 10^{-2}}_{a.u.}
\end{equation*} 
We require maximization of the emission in the neighbourhood of the second harmonic:
\begin{equation*}%\label{eq:HCl02har}
	\omega_{2,0} = E_2 - E_0 = {2.54\cdot 10^{-2}}_{a.u.}
\end{equation*}
$\tfeps$ and $\tfmu$ are chosen to be rectangular functions.

In order to prevent dissociation of the molecule, the energies above the dissociation threshold were restricted (see Section~\ref{sssec:dis}).

The details of the problem are summarised in Table~\ref{tab:HCl}.

\begin{table}
\begin{equation*}
	\renewcommand{\arraystretch}{2}
	\begin{array}{|c||c|}
		\hline
		\operator{H}_0 & \frac{\operator{P}^2}{2\cdot 1785} + 0.171\left[\exp\left(-0.975\,\operator{X}\right)-\operator{I}\right]^2 \\ \hline
		\operator{\mu} &  \left(0.19309\,\operator{X}\right)\;\times \\
		& \left\lbrace \operator{I}-\Real\left[\tanh\left((0.17069 +0.056854\,i)\left(\operator{X} - 0.10630\,\operator{I}\right)^{1.8977}\right)\right]\right\rbrace\\ \hline
		\ket{\psi_0} & \ket{\varphi_0} \\ \hline
		T & 10^4 \\ \hline
		\tfeps & 2500\,\Theta(0.015-\omega) \\ \hline
		\tfmu & 100\,\Theta(\omega - 0.025)\,\Theta(0.027-\omega) \\ \hline
		L & 19 \\ \hline
		\gamma_n & 
			\left\lbrace
			\begin{smallmatrix}
				0 & \quad n\leq 19 \\
				(n-19)^2 & \quad n>19
			\end{smallmatrix}			
			\right. \\ \hline
		\bar\epsilon^{0}(\omega) & \Theta(0.015-\omega) \\ \hline
		K_i & 1 \\ \hline
		x \text{ domain} & [-0.69407,\;3.51178) \\ \hline
		N_{x} & 32 \\ \hline
		\tau & 10^{-3} \\ \hline
	\end{array}
\end{equation*}
\caption{The details of the HCl problem}\label{tab:HCl}
\end{table}

\begin{figure}
	\centering \includegraphics[width=3in]{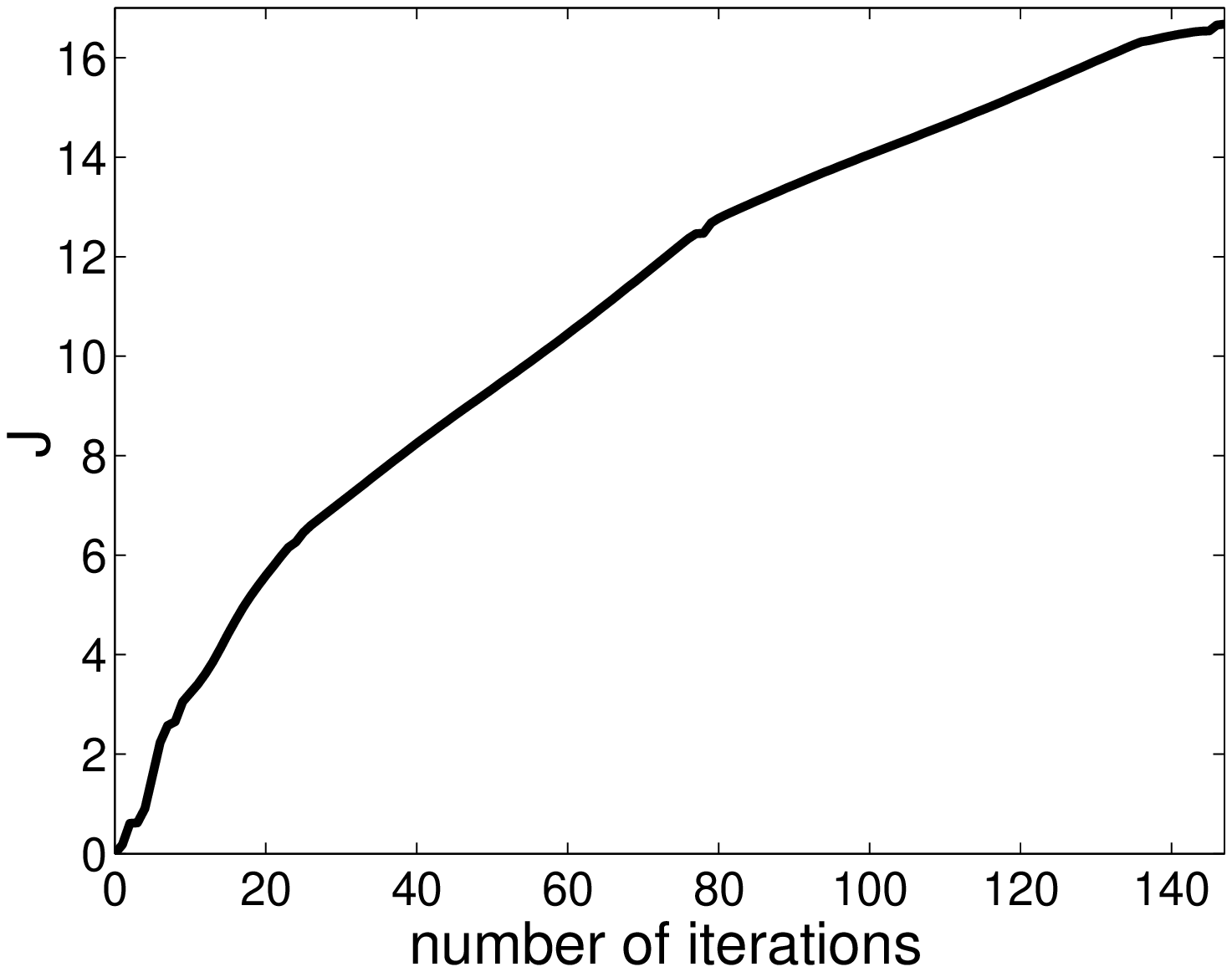}
	\caption{The convergence curve of the HCl problem}\label{fig:HClconv}	
\end{figure}

\begin{figure}
	\centering \includegraphics[width=3in]{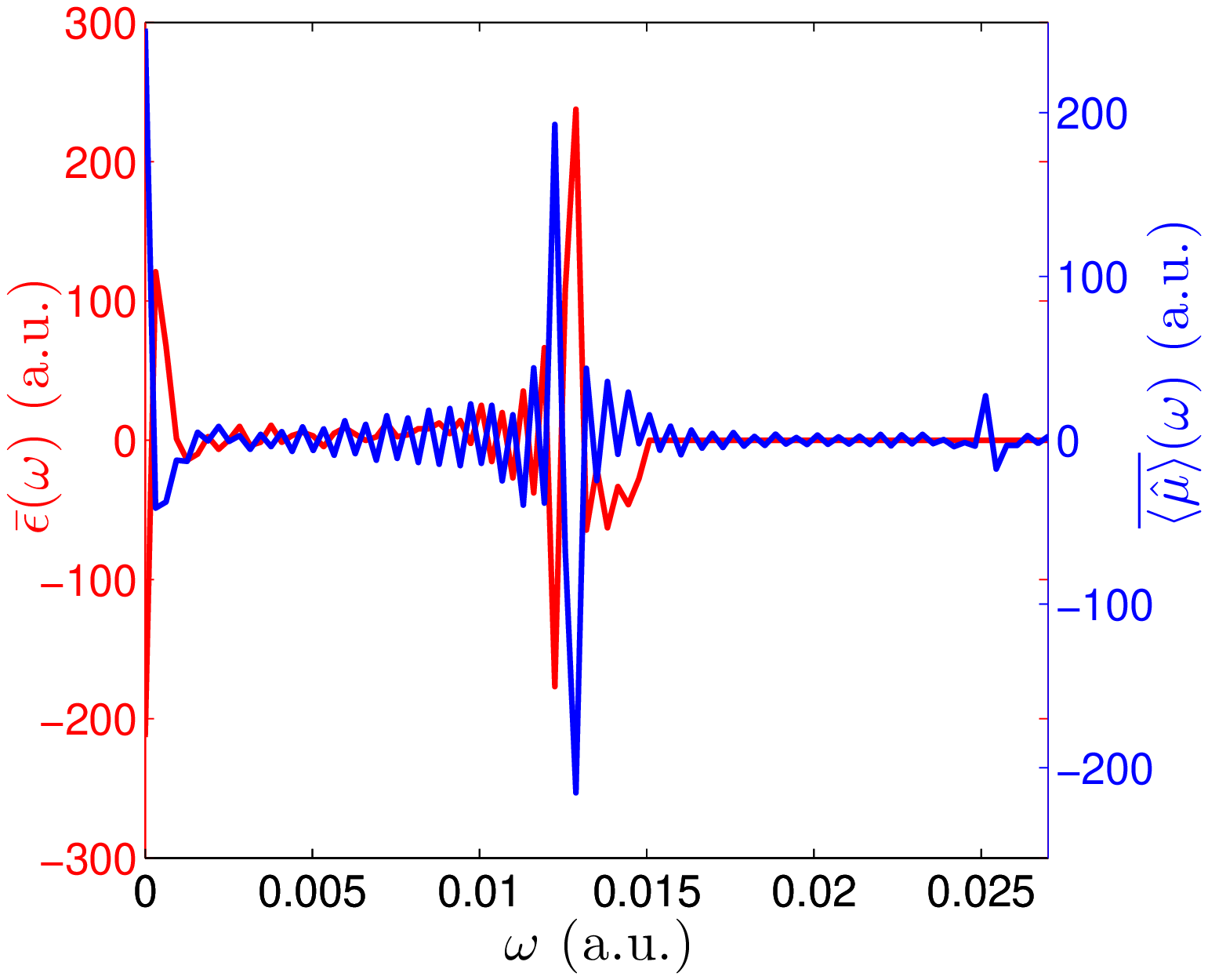}
	\caption{(Color online) The spectra of the driving field (red, gray) and the dipole expectation spectrum (blue, dark gray), for the HCl problem. Two significant peaks appear: A large linear response of the dipole to the driving field, 
	and a significant non-linear response in the neighbourhood of the second harmonic.
	Notice the low frequency component of the driving field which changes the static part of the Hamiltonian.}\label{fig:HClepsmu}	
\end{figure}

The convergence curve is shown in Fig.~\ref{fig:HClconv}. The resulting $\epsw$ and $\muw$ are shown in Fig.~\ref{fig:HClepsmu}. 
$\muw$ mainly consists of a large linear response to the driving field, as could be expected. 
However, there is also a significant non-linear response in the region of the second harmonic frequency, as required.

A detailed analysis of the results of the first three examples may be found in~\cite[Chapter~4]{thesis}.

\subsection{One dimensional particle in a truncated Coulomb potential}\label{ssec:coulomb}

The last example demonstrates the application of the new method in a system with stronger non-linearity. A driven electron in a Coulomb potential is the system studied. The model
has been extensively studied in the context of harmonic generation \cite{lewenstein94,agostini99}.
Typically, for strong driving fields a comb of odd frequencies is generated up to a cutoff.
In the present example the target is the emission of a single high harmonic of the driving frequency.
This target has similarities to the experiment in JILA \cite{murnane04} where the emission of the 27'th harmonic was enhanced relative to its neighbours using a genetic algorithm optimization.

\begin{figure}
	\centering \includegraphics[width=3in]{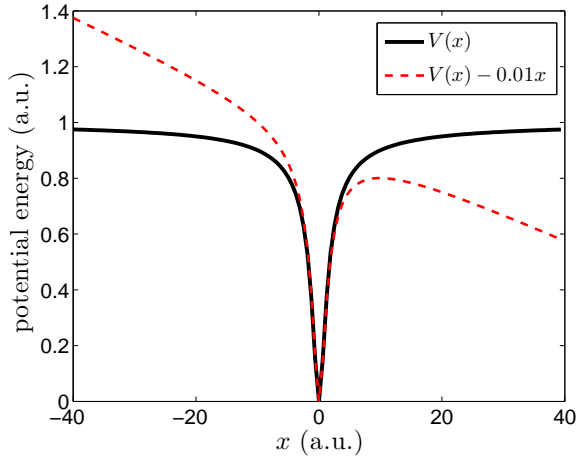}
	\caption{(Color online) The truncated Coulomb potential (Eq.~\eqref{eq:coulV}, solid black), and the potential energy of the system under the influence of a strong field (dashed red).}\label{fig:coulV}	
\end{figure}

Our model consists of a particle of unit mass and charge placed in a truncated Coulomb potential 
constrained to one dimension (see Fig.~\ref{fig:coulV}):
\begin{equation}\label{eq:coulV}
	V(x) = 1 - \frac{1}{\sqrt{x^2 + 1}}
\end{equation}
The dipole operator is \text{$\operator{\mu} = \operator{X}$}.

\begin{figure}
	\centering \includegraphics[width=3in]{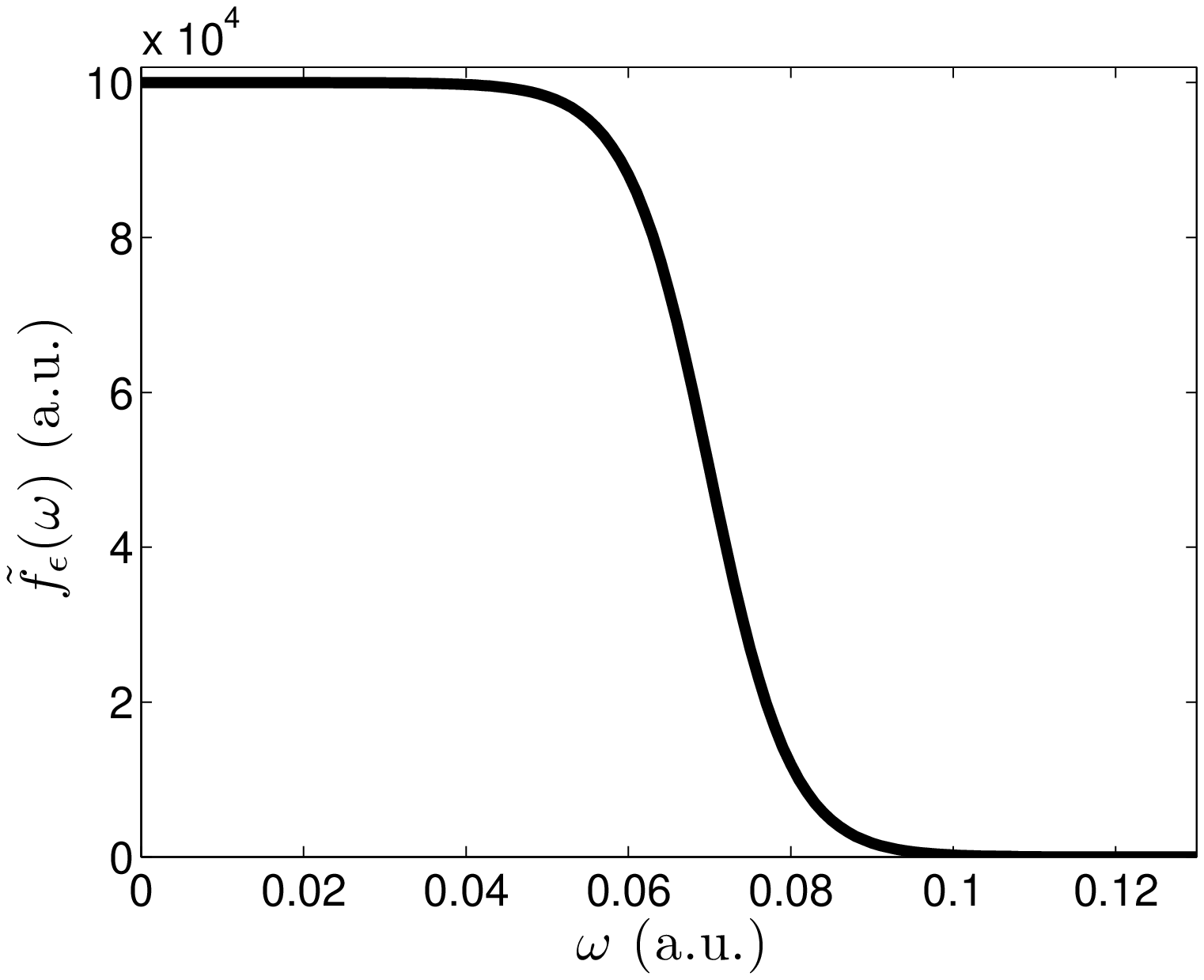}
	\caption{$\tfeps$ of the truncated Coulomb potential problem}\label{fig:coultfeps}	
\end{figure}

The driving field is restricted to frequencies which are much lower than the resonance frequencies of the system. $\epsw$ is restricted so as not to exceed the region of \text{$\omega=0.07_{a.u.}$}. The shape of $\tfeps$ (see Fig.~\ref{fig:coultfeps}) induces a smooth filtration of higher frequencies. We require maximization of the emission in the region of one of the Bohr frequencies of the system, \text{$\omega_{5,0}=0.624_{a.u.}$}. For the filter $\tfmu$ a rectangular function is employed.

\begin{figure}
	\centering \includegraphics[width=3in]{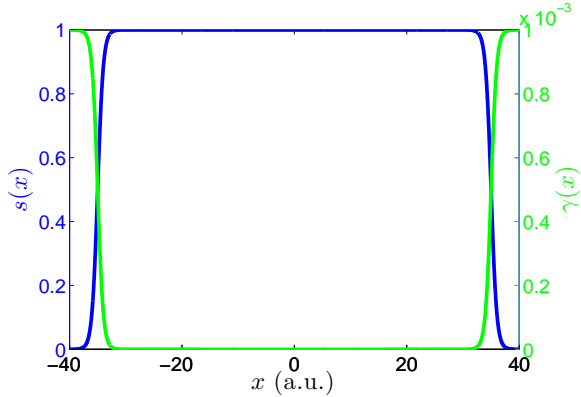}
	\caption{(Color online) $s(x)$ (blue, dark gray) and $\gamma(x)$ (green, light gray) of the truncated Coulomb potential problem}\label{fig:sgamma}	
\end{figure}

The edges of the $x$ grid are restricted using the method presented in Section~\ref{sssec:dis}. $s(x)$ and $\gamma(x)$ are shown in Fig.~\ref{fig:sgamma}.

The details of the problem are summarised in Table~\ref{tab:coul}.

\begin{table}
\begin{equation*}
	\renewcommand{\arraystretch}{1.5}
	\begin{array}{|c||c|}
		\hline
		\operator{H}_0 & \frac{\operator{P}^2}{2} + \operator{I} - \frac{1}{\sqrt{\operator{X}^2 + \operator{I}}} \\ \hline
		\operator{\mu} &  \operator{X} \\ \hline
		\ket{\psi_0} & \ket{\varphi_0} \\ \hline
		T & 2000 \\ \hline
		\tfeps & 5\times 10^4\,\lbrace 1-\tanh[100(\omega - 0.07)]\rbrace \\ \hline
		\tfmu & \Theta(\omega - 0.61)\,\Theta(0.63-\omega) \\ \hline
		s(x) & 0.5\,[\tanh(x + 35) - \tanh(x - 35)] \\ \hline
		\gamma(x) & 10^{-3}\,[1 - s(x)] \\ \hline
		\bar\epsilon^{0}(\omega) &  0.3\,\lbrace 1-\tanh[100(\omega - 0.07)]\rbrace \\ \hline
		K_i & 10^{-6} \\ \hline
		x \text{ domain} & [-40,\;40) \\ \hline
		N_{x} & 128 \\ \hline
		\tau & 5\times 10^{-4} \\ \hline
	\end{array}
\end{equation*}
\caption{The details of the truncated Coulomb potential problem}\label{tab:coul}
\end{table}

\begin{figure}
	\centering \includegraphics[width=3in]{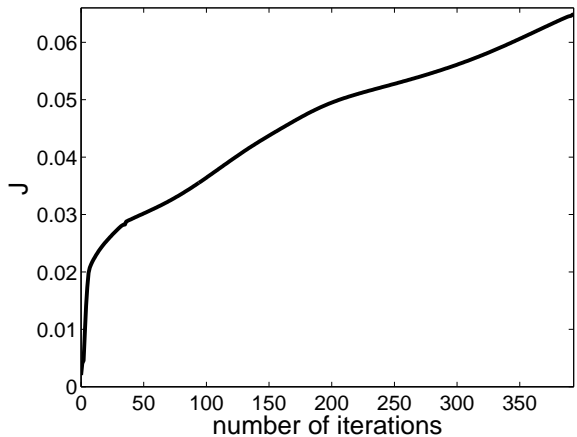}
	\caption{The convergence curve of the truncated Coulomb potential problem}\label{fig:coulconv}	
\end{figure}

The convergence curve is shown in Fig.~\ref{fig:coulconv}.

\begin{figure}
	\centering \includegraphics[width=3in]{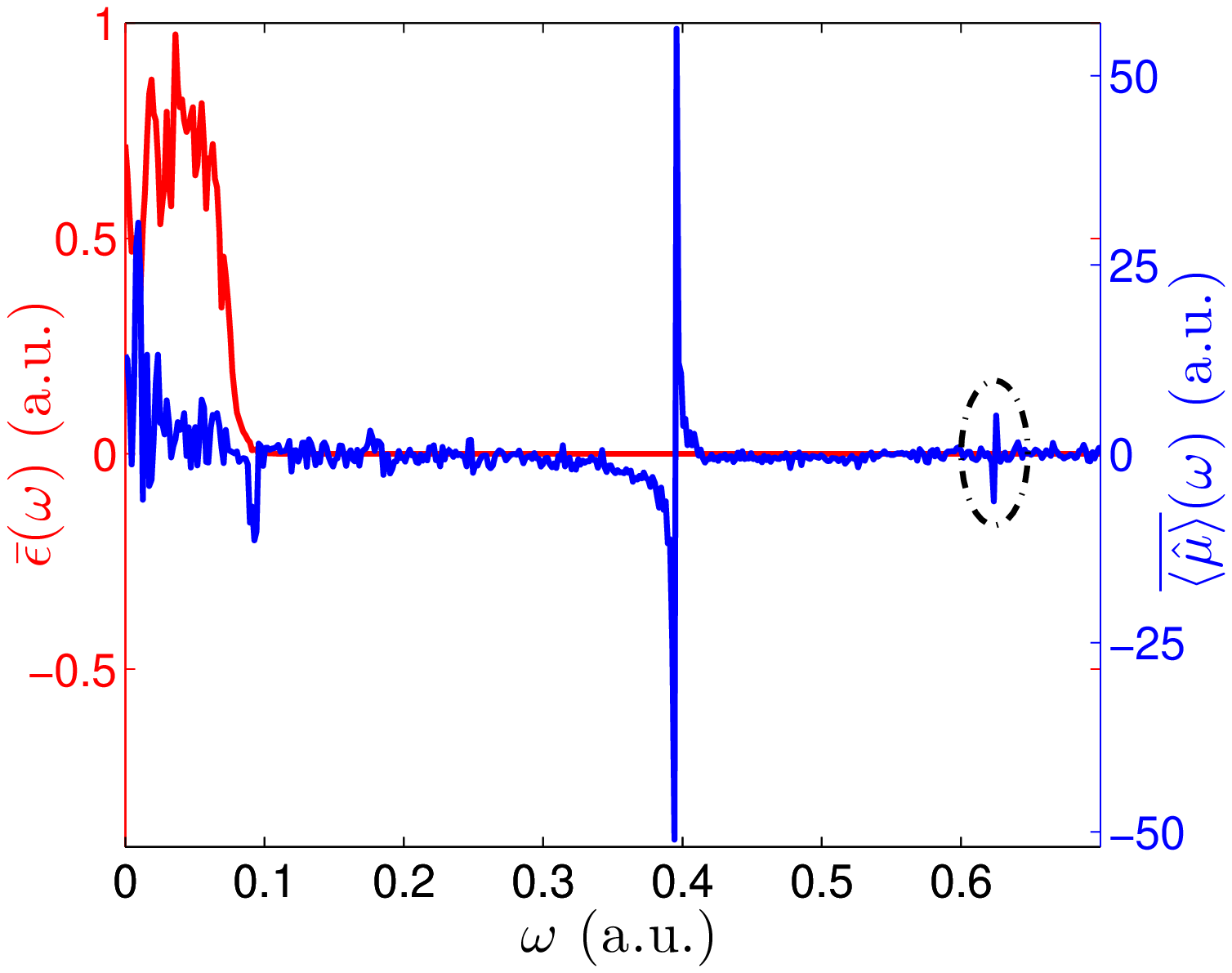}
	\caption{(Color online) The spectra of the driving field (red, gray) and the dipole expectation spectrum (blue, dark gray), for the truncated Coulomb potential problem; the response in the desired frequency, \text{$\omega_{5,0}=0.624_{a.u.}$}, is marked by a black ellipse. The largest peak of $\muw$ is in the fundamental frequency, \text{$\omega_{1,0}=0.395_{a.u.}$}.}\label{fig:coulepsmu}	
\end{figure}

The resulting field $\epsw$ and dipole 
$\muw$ are shown in Fig.~\ref{fig:coulepsmu}. The largest peak of $\muw$ is located near the fundamental frequency of the system, \text{$\omega_{1,0}=0.395_{a.u.}$}. The response in the desired frequency is marked on the figure. The method is shown to be effective also in the production of frequencies considerably higher than those of the driving field.

The final solution obtained up to the chosen $\tau$ is not a converged solution, as can be deduced from the convergence curve shape at the end of the optimization process (see Fig.~\ref{fig:coulconv}). If the optimization process is carried on to smaller $\tau$, contributions to $\muw$ from interactions with the boundaries of the $x$ grid begin to be significant. Nevertheless, a converged solution which does not involve such undesirable effects can be obtained. 
This is achieved by the increment of the $\gamma(x)$ values during the optimization process 
before the spurious effects appear.

Maximizing the response of an arbitrary frequency, which is not one of the Bohr frequencies, was also achieved. %However, the resulting amplitude of the response was found to be considerably smaller than if a Bohr frequency was selected.
However, the resulting amplitude of the response was found to be considerably smaller than the response for a Bohr frequency. Larger response was obtained 
when partial ionization was allowed. Technically, this requires absorbing boundary conditions. This topic is still under investigation, and is therefore not presented. 

\section{Conclusion}\label{sec:con}

Optimizing harmonic generation is one of the most difficult tasks in the context of
quantum control. A major obstacle is that the target objective cannot be formulated in the time domain.
Additional restrictions have to be added to suppress ionization or dissociation. 
A new theoretical method of calculation for optimal control of harmonic generation was studied. The task was addressed by the means of OCT using a frequency domain formulation. The relaxation method was used as the iterative optimization procedure. 

For low and intermediate control fields fast convergence was obtained when the emitted 
high harmonic fit a fundamental Bohr frequency of the system. Stronger fields can modify the system
thus allowing harmonic emission in other frequencies. The difficulty we found in 
locating such solutions was that they competed with dissociation or ionization. Thus, the method should be modified to allow partial dissociation or ionization.

This paper focuses on the control aspect of the new method. However, the physical interpretation of the results is of interest. 
Significant  physical insight can be deduced from the optimized fields, unravelling new harmonic generation mechanisms, 
as was demonstrated in~\cite[Chapter~4]{thesis}. 
Further studies employing the current approach will contribute to the understanding of harmonic generation processes, 
in particular non-adiabatic mechanisms which go beyond the three-step-model \cite{corkum93}.

\begin{acknowledgments}
We thank Christiana Koch, Hardy Gross and Nimrod Moiseyev for helpful discussions and criticism.
We gratefully acknowledge financial support from 
the Israel Science Foundation.
The Fritz Haber
Center is supported by the Minerva Gesellschaft f\"{u}r die Forschung
GmbH M\"{u}nchen, Germany. 
\end{acknowledgments}

\appendix

\section{The derivation of the Euler-Lagrange equations}\label{app:der}
%The derivation is performed for the general case, with the exception  that \text{$\commut{\operator{\mu}}{\operator{O}} \neq \operator{0}$} and \text{$\kappa>0$}.
The general maximization functional is rewritten in its full form, for convenience:
\begin{align}
	& J \equiv J_{max} + J_{bound} + J_{forb} + J_{penal} + J_{con} \label{eq:Jwd}\\ 
	& J_{max} \equiv \frac{1}{2}\int_0^\Omega \tfO\overline{\exval{\operator{O}_a}}^2(\omega)\,d\omega  & \tfO\geq 0 \label{eq:Jmaxd} \\
%	& \operator{O}_a \equiv \operator{P}_{a}\operator O\operator{P}_{a} \label{eq:Oad} \\
%	& \operator{P}_{a} \equiv \sum_{n=0}^L\ket{\varphi_n}\bra{\varphi_n} \label{eq:Pad} \\
	& \Oaw \equiv \sqrt{\frac{2}{\pi}}\int_0^T \exval{\operator{O}_a}(t)\cos(\omega t)\,dt \label{eq:Oawd} \\
	& J_{bound} \equiv -\frac{1}{2}\kappa\left[\deriv{\exval{\operator{O}}(T)}{t}\right]^2 & \kappa \geq 0 \label{eq:Jboundd}\\
	& J_{forb} \equiv -\int_0^T \bracketsO{\psi(t)}{\operator{P}_f^\gamma}{\psi(t)}\,dt \label{eq:Jforbd}\\
%	& \operator{P}_f^\gamma \equiv \sum_{n=L+1}^{N-1}\gamma_n\ket{\varphi_n}\bra{\varphi_n} & \gamma_n>0 \label{eq:Pfd} %\\
%\end{align}
%\begin{align}
	& J_{penal} \equiv -\int_0^\Omega\frac{1}{\tfeps}\bar{\epsilon}^2(\omega)\,d\omega & \tfeps>0 \label{eq:Jpenald}\\
	& \epsw \equiv \sqrt{\frac{2}{\pi}}\int_0^T \epsilon(t)\cos(\omega t)\,dt \label{eq:epswd} \\
	& J_{con} \equiv -2\Real{\int_0^T\bracketsO{\chi(t)}{\pderiv{}{t}+i\operator H(t)}{\psi(t)}\,dt} \label{eq:Jcond}\\
	& \operator{H}(t) = \operator{H}_0 - \operator{\mu}\epsilon(t) = \operator{H}_0 - \operator{\mu}\left(\sqrt{\frac{2}{\pi}}\int_0^\Omega \bar{\epsilon}(\omega)\cos(\omega t)\,d\omega\right)\label{eq:Hd}
\end{align}

The constraint equations are:
\begin{align}
	\pderiv{\ket{\psi(t)}}{t} &= -i\operator{H}(t)\ket{\psi(t)} & & \ket{\psi(0)} = \ket{\psi_0}
	\label{eq:Schrd}\\
	\pderiv{\bra{\psi(t)}}{t} &= i\bra{\psi(t)}\operator{H}(t) & & \bra{\psi(0)} = \bra{\psi_0} \label{eq:conjSchrd}
\end{align}
Eqs.~\eqref{eq:Schrd}, \eqref{eq:conjSchrd} ensure that:
\[
	\bra{\psi(t)} = \ket{\psi(t)}^+
\]
Assuming this, all the computations can be performed using \eqref{eq:Schrd} only.

The extremum conditions are:
\begin{align}
	&\fnlderiv{J}{\bar{\epsilon}(\omega)} = 0 \label{eq:dJdepswd}\\
	&\fnlderiv{J}{\ket{\psi(t)}} = 0 \label{eq:dJdpsitkd} \\ 
	&\fnlderiv{J}{\bra{\psi(t)}} = 0 \label{eq:dJdpsitbd} \\	
	&\fnlderiv{J}{\ket{\psi(T)}} = 0 \label{eq:dJdpsiTkd} \\ 
	&\fnlderiv{J}{\bra{\psi(T)}} = 0 \label{eq:dJdpsiTbd}
\end{align}

After integrating by parts the following expression in $J_{con}$:
\[
	\int_0^T\bracketsbiggm{\chi(t)}{\pderiv{\psi(t)}{t}}\,dt
\]
we obtain:
\begin{align}
	J_{con} = -2\Real&\left[\bracketsbiggm{\chi(T)}{\psi(T)}-\bracketsbiggm{\chi(0)}{\psi(0)} - \int_0^T\bracketsbiggm{\left(\pderiv{}{t}+i\operator H(t)\right)\chi(t)}{\psi(t)}\,dt\right]
%	J_{con} = -2\Real&\left[\bracketsbiggm{\chi(T)}{\psi(T)}-\bracketsbiggm{\chi(0)}{\psi(0)}\right. \nonumber \\
%	  &\left.-\int_0^T\bracketsbiggm{\left(\pderiv{}{t}+i\operator H(t)\right)\chi(t)}{\psi(t)}\,dt\right]
	  \label{eq:Jconp}
\end{align}

For simplicity, we will assume that $J_{bound}$ has no explicit dependence on $\epsw$. This requires that \text{$\commut{\operator{\mu}}{\operator{O}}=\operator{0}$} (see Eq.~\eqref{eq:Jbounduse}), or that \text{$\kappa=0$}. The expression for the LHS of \eqref{eq:dJdepswd} is obtained using \eqref{eq:Jpenald}, \eqref{eq:Jconp}, \eqref{eq:Hd}:
\begin{align}
	\fnlderiv{J}{\epsw} =& \fnlderiv{J_{penal}}{\epsw} + \fnlderiv{J_{con}}{\epsw} \label{eq:dJdepscomp}\\
	\fnlderiv{J_{penal}}{\epsw} =& -\frac{2}{\tfeps}\epsw \label{eq:dJpenaldeps}\\
	\fnlderiv{J_{con}}{\epsw} =& 2\,\Real\left[-i\int_0^T\bracketsO{\chi(t)}{\fnlderiv{\operator{H}(t)}{\epsw}}{\psi(t)}\,dt\right] \nonumber \\
	=&-2\,\Imag\left[\sqrt{\frac{2}{\pi}}\int_0^T\bracketsO{\chi(t)}{\operator{\mu}}{\psi(t)}\cos(\omega t)\,dt\right] \nonumber \\
	=&-2\,\Imag\left\lbrace \mathcal{C}\left[\bracketsO{\chi(t)}{\operator{\mu}}{\psi(t)}\right]\right\rbrace \label{eq:dJcondeps}
\end{align}
From \eqref{eq:dJdepswd}, \eqref{eq:dJdepscomp}, \eqref{eq:dJpenaldeps}, \eqref{eq:dJcondeps}, we obtain the following expression for $\epsw$:
\begin{equation}\label{eq:epswresult}
	\epsw = \tfeps\mathcal{C}\left[-\Imag{\bracketsO{\chi(t)}{\operator{\mu}}{\psi(t)}}\right] 
\end{equation}

In order to derive the LHS of \eqref{eq:dJdpsitkd}, we first write the explicit expression of $J_{max}$ as a functional of $\ket{\psi(t)}$:
\begin{align}
	&J_{max} = \frac{1}{\pi}\!\int_0^\Omega\!\int_0^T\!\int_0^T \tfO \bracketsO{\psi(t)}{\operator{O}_a}{\psi(t)} \bracketsO{\psi(t')}{\operator{O}_a}{\psi(t')}\cos(\omega t)\cos(\omega t')\,dt\,dt'\,d\omega 
%	&J_{max}= \nonumber \\
%	&\frac{1}{\pi}\!\int_0^\Omega\!\int_0^T\!\int_0^T \tfO \bracketsO{\psi(t)}{\operator{O}_a}{\psi(t)} \bracketsO{\psi(t')}{\operator{O}_a}{\psi(t')}\cos(\omega t)\cos(\omega t')\,dt\,dt'\,d\omega 
	\label{eq:Jmaxex}
\end{align}
The expression for the LHS of \eqref{eq:dJdpsitkd} is obtained using \eqref{eq:Jmaxex}, \eqref{eq:Jforbd}, \eqref{eq:Jconp}:
\begin{align}
	\fnlderiv{J}{\ket{\psi(t)}} =& \fnlderiv{J_{max}}{\ket{\psi(t)}} + \fnlderiv{J_{forb}}{\ket{\psi(t)}} + \fnlderiv{J_{con}}{\ket{\psi(t)}} \label{eq:dJdpsicomp}\\
	\fnlderiv{J_{max}}{\ket{\psi(t)}} =& \frac{2}{\pi}\int_0^\Omega\!\int_0^T \tfO \bra{\psi(t)}\operator{O}_a \bracketsO{\psi(t')}{\operator{O}_a}{\psi(t')}\cos(\omega t)\cos(\omega t')\,dt'\,d\omega\nonumber \\
	= &\sqrt{\frac{2}{\pi}}\int_0^\Omega \tfO\Oaw\cos(\omega t)\,d\omega\bra{\psi(t)}\operator{O}_a \nonumber \\
	= &\mathcal{C}^{-1}\left[\tfO\Oaw\right]\bra{\psi(t)}\operator{O}_a \label{eq:dJmaxdpsi}\\
	\fnlderiv{J_{forb}}{\ket{\psi(t)}} = & -\bra{\psi(t)}\operator{P}_f^\gamma \label{eq:dJforbdpsi}\\
	\fnlderiv{J_{con}}{\ket{\psi(t)}} = & \pderiv{\bra{\chi(t)}}{t} + \bra{i\operator{H}(t)\chi(t)} \label{eq:dJcondpsi}
\end{align}
Using \eqref{eq:dJdpsitkd}, \eqref{eq:dJdpsicomp}, \eqref{eq:dJmaxdpsi}, \eqref{eq:dJforbdpsi}, \eqref{eq:dJcondpsi}, we obtain:
\begin{equation}\label{eq:ihSchrbra}
	\pderiv{\bra{\chi(t)}}{t} =  -\bra{i\operator{H}(t)\chi(t)} - \bra{\psi(t)} \left\lbrace\mathcal{C}^{-1}\left[\tfO\Oaw\right]\operator{O}_a - \operator{P}_f^\gamma\right\rbrace
\end{equation}
Eq.~\eqref{eq:dJdpsitbd} gives the adjoint of \eqref{eq:ihSchrbra}:
\begin{equation}\label{eq:ihSchrket}
	\pderiv{\ket{\chi(t)}}{t} = -i\operator{H}(t)\ket{\chi(t)} - \left\lbrace\mathcal{C}^{-1}\left[\tfO\overline{\exval{\operator{O}_a}}(\omega)\right]\operator{O}_a - \operator{P}_f^\gamma\right\rbrace\ket{\psi(t)}
\end{equation}

In order to derive the expression of the LHS of \eqref{eq:dJdpsiTkd}, we write \eqref{eq:Jboundd} in a more useful form. Taking the expectation value of both sides of the Heisenberg equation, we have:
\begin{equation}\label{eq:Heisd}
	\deriv{\exval{\operator{O}}(T)}{t} = i\exval{\left[\operator{H}(T), \operator{O}\right]}(T)
\end{equation}
In the special case that \text{$\commut{\operator{\mu}}{\operator{O}}=\operator{0}$}, we have:
\begin{equation}\label{eq:Heis0d}
	\deriv{\exval{\operator{O}}(T)}{t} = i\exval{\left[\operator{H}_0, \operator{O}\right]}(T)
\end{equation}
In this case, $J_{bound}$ becomes:
\begin{equation}\label{eq:Jbounduse}
	J_{bound} = \frac{\kappa}{2}\bracketsO{\psi(T)}{\left[\operator{H}_0, \operator{O}\right]}{\psi(T)}^2
\end{equation}
The LHS of \eqref{eq:dJdpsiTkd} is:
\begin{align}
	&\fnlderiv{J}{\ket{\psi(T)}}=\fnlderiv{J_{bound}}{\ket{\psi(T)}} + \fnlderiv{J_{con}}{\ket{\psi(T)}} \label{eq:dJdpsiTcomp} \\
	&\fnlderiv{J_{bound}}{\ket{\psi(T)}} = \kappa\bracketsO{\psi(T)}{\left[\operator{H}_0, \operator{O}\right]}{\psi(T)}\bra{\psi(T)}\left[\operator{H}_0, \operator{O}\right] \label{eq:dJbounddpsiT} \\
	&\fnlderiv{J_{con}}{\ket{\psi(T)}} = -\bra{\chi(T)} \label{eq:dJcondpsiT}
\end{align}
Using \eqref{eq:dJdpsiTkd}, \eqref{eq:dJdpsiTcomp}, \eqref{eq:dJbounddpsiT}, \eqref{eq:dJcondpsiT}, we obtain:
\begin{align}
	\bra{\chi(T)} =& \kappa\bracketsO{\psi(T)}{\left[\operator{H}_0, \operator{O}\right]}{\psi(T)}\bra{\psi(T)}\left[\operator{H}_0, \operator{O}\right] \nonumber \\
	=& \kappa\exval{\left[\operator{H}_0, \operator{O}\right]}(T)\bra{\psi(T)}\left[\operator{H}_0, \operator{O}\right] \label{eq:chiTbra}
\end{align}
Eq.~\eqref{eq:dJdpsiTbd} gives the adjoint of \eqref{eq:chiTbra}:
\begin{equation} \label{eq:chiTket}
	\ket{\chi(T)} = \kappa\exval{\left[\operator{H}_0, \operator{O}\right]}(T) \left[\operator{H}_0, \operator{O}\right]\ket{\psi(T)}
\end{equation}
Eqs.~\eqref{eq:ihSchrbra}, \eqref{eq:ihSchrket}, \eqref{eq:chiTbra}, \eqref{eq:chiTket} ensure that:
\[
	\bra{\chi(t)} = \ket{\chi(t)}^+
\]
Assuming this, all the computations can be performed using \eqref{eq:ihSchrket}, \eqref{eq:chiTket} only.

We collect the resulting equations, \eqref{eq:epswresult}, \eqref{eq:ihSchrket}, \eqref{eq:chiTket}, together with the constraint \eqref{eq:Schrd}:
\begin{align}
	&\pderiv{\ket{\psi(t)}}{t} = -i\operator{H}(t)\ket{\psi(t)}, \nonumber\\
	& \hspace{2cm}\ket{\psi(0)} = \ket{\psi_0} \\ 
	&\pderiv{\ket{\chi(t)}}{t} = -i\operator{H}(t)\ket{\chi(t)} - \left\lbrace\mathcal{C}^{-1}\left[\tfO\overline{\exval{\operator{O}_a}}(\omega)\right]\operator{O}_a - \operator{P}_f^\gamma\right\rbrace\ket{\psi(t)}, \nonumber\\
	&\hspace{2cm} \ket{\chi(T)} = \kappa\exval{\left[\operator{H}_0, \operator{O}\right]}(T) \left[\operator{H}_0, \operator{O}\right]\ket{\psi(T)} \\
	& \operator{H}(t) = \operator{H}_0 - \operator{\mu}\epsilon(t) \nonumber \\
	& \epsw = \tilde{f}_{\epsilon}(\omega)\mathcal{C}\left[-\Imag{\bracketsO{\chi(t)}{\operator{\mu}}{\psi(t)}}\right] \\
	& \epsilon(t) = \mathcal{C}^{-1}[\epsw]
\end{align}
These are the Euler-Lagrange equations of the problem.

\section{The approximation of the HCl molecule}\label{app:HCl}

The potential of the H-Cl bond was obtained by adjusting the parameters of the Morse potential
\begin{equation}\label{eq:Morse}
	V(x) = D_0\left[\exp(-ax)-1\right]^2
\end{equation}
to experimental data on HCl --- The atomization energy of HCl, and the frequency of vibration, using the IR absorption frequency for the transition to the fundamental state. We made a few reasonable approximations. The resulting potential is presented in Table~\ref{tab:HCl} (the second term of $\operator{H}_0$).
 
The dipole function was obtained by adjusting experimental data to a reasonable functional form. The experimental data is the first 4 derivatives of $\mu(x)$ at equilibrium \cite{HCl}:
\[
	\left(\nderiv{\mu}{x}{n}\right)_{eq} \qquad\qquad n=1,2,3,4
\]
The functional form is:
\begin{equation}\label{eq:mutanh}
	\mu(x) = a_1x\left\lbrace 1-\tanh\left[a_2(x-a_3)^{a_4}\right]\right\rbrace
\end{equation}
%This functional form is intended to represent a nearly linear form at $x=0$, that decays to $0$ at the dissociation region of the potential. This behaviour is typical to a homolitic bond cleavage.
We made the approximation:
\[
	\left(\nderiv{\mu}{x}{n}\right)_{x=0}\approx\left(\nderiv{\mu}{x}{n}\right)_{eq}
\]
The resulting system of equations was solved using the Symbolic Math Toolbox of MATLAB\@. The resulting function is complex. We take its real part (see Table~\ref{tab:HCl}).

%\bibliography{harminicgeneration,../Database/pub}
\bibliography{harminicgeneration}

\end{document}